\documentclass
[prd,amsfonts,amssymb,twocolumn,a4paper,balancelastpage,twoside]{revtex4}%
\usepackage{amsfonts}
\usepackage{amsmath}
\usepackage{amssymb}
\usepackage{graphicx}%
\providecommand{\U}[1]{\protect\rule{.1in}{.1in}}

\begin{document}
\preprint{ }
\title[ ]{Massive cosmic strings in Bianchi type II}
\author{Jos\'e Antonio Belinch\'on}
\affiliation{Dept. F\'isica. ETS Arquitectura. UPM. Av. Juan de Herrera $4$. Madrid
$28040$. Espa\~{n}a}

\begin{abstract}
We study a massive cosmic strings with BII symmetries cosmological models in
two contexts. The first of them is the standard one with a barotropic equation
of state. In the second one we explore the possibility of taking into account
variable \textquotedblleft constants\textquotedblright\ ($G$ and $\Lambda
).$Both models are studied under the self-similar hypothesis. We put special
emphasis in calculating the numerical values for the equations of state. We
find that for $\omega\in(0,1]$, $G$, is a growing time function while
$\Lambda$, behaves as positive decreasing time function. If $\omega=0,$ both
\textquotedblleft constants\textquotedblright, $G$ and $\Lambda,$ behave as
true constants.

\end{abstract}
\date{\today}
\maketitle

\section{Introduction}

The exponential expansion of the Universe (inflationary era) causes the
Universe to heat up to a very high temperature so the subsequent evolution of
the Universe is exactly as in hot BB model. Hence, the phase transition (as
the temperature falls below some critical temperature) in the early universe
causes topologically stable defects: vacuum domain walls, strings and
monopoles (see \cite{Z1} and \cite{v1}). But domain walls and monopoles are
disastrous for the cosmological models. Strings, on the other hand, causes no
harm, but can lead to very interesting astrophysical consequences (see
\cite{k}). Also the existence of a large scale network of strings the early
universe does not contradict the present-day observations. The vacuum strings
may generate density fluctuations sufficient to explain the galaxy formation
(see \cite{Z2}).

The relativistic treatment of strings was initiated by Letelier (see
\cite{L1}-\cite{L2}) and Stachel (see \cite{S}). Here we have considered
gravitational effects, arised from strings by coupling of stress energy of
strings to the gravitational field. Letelier (see \cite{L2}) defined the
massive strings as the geometric strings (massless) with particles attached
along its expansions.

The strings that form the cloud are massive strings instead of geometrical
strings. Each massive string is formed by a geometrical string with particles
attached along its extension. Hence, the string that form the cloud are the
generalization of Takabayasi's relativistic model of strings (called
$p-$strings). This is simplest model wherein we have particles and strings
together. In principle we can eliminate the strings and end up with a cloud of
particles. This a desirable property of a model of a string cloud to be used
in cosmology since strings are not observed at the present time of evolution
of the universe (see \cite{B}-\cite{saha1}).

In modern cosmological theories, the cosmological constant remains a focal
point of interest (see \cite{cc1}-\cite{cc4}\ \ for reviews of the problem). A
wide range of observations now compellingly suggest that the universe
possesses a non-zero cosmological constant. Some of the recent discussions on
the cosmological constant \textquotedblleft problem\textquotedblright\ and on
cosmology with a time-varying cosmological constant point out that in the
absence of any interaction with matter or radiation, the cosmological constant
remains a \textquotedblleft constant\textquotedblright. However, in the
presence of interactions with matter or radiation, a solution of Einstein
equations and the assumed equation of covariant conservation of stress-energy
with a time-varying $\Lambda$ can be found. This entails that energy has to be
conserved by a decrease in the energy density of the vacuum component followed
by a corresponding increase in the energy density of matter or radiation
\ Recent observations strongly favour a significant and a positive value of
$\Lambda$ with magnitude $\Lambda(G\hbar/c^{3})\approx10^{-123}$. These
observations suggest on accelerating expansion of the universe, $q<0$.

Our current understanding of the physical universe is anchored on the analysis
of expanding, isotropic and homogeneous models with a cosmological constant,
and linear perturbations thereof. This model successfully accounts for the
late time universe, as is evidenced by the observation of large scale cosmic
microwave background observations. Parameter determination from the analysis
of CMB fluctuations appears to confirm this picture. However, further analyses
seem to suggests some inconsistency. In particular, its appears that the
universe could have a preferred direction. Followup analyses of various sets
of WMAP data sets, with different techniques seem to lead to the same
conclusion. It is still unclear whether or not the directional preference is
intrinsic to the underlying model, and what implications this has on our
understanding of cosmology. For this reason Bianchi models are important in
the study of anisotropies.

The study of SS models is quite important since a large class of orthogonal
spatially homogeneous models are asymptotically self-similar at the initial
singularity and are approximated by exact perfect fluid or vacuum self-similar
power law models. Exact self-similar power-law models can also approximate
general Bianchi models at intermediate stages of their evolution. This last
point is of particular importance in relating Bianchi models to the real
Universe. At the same time, self-similar solutions can describe the behaviour
of Bianchi models at late times i.e. as $t\rightarrow\infty$ (see
\cite{ColeyDS})$.$A particular interest is in determining the exact value of
the equations of state (see \cite{Wainwrit}). The geometry and physics at
different points on an integral curve of a homothetic vector field (HVF)
differ only by a change in the overall length scale and in particular any
dimensionless scalar will be constant along the integral curves. In this sense
the existence of a HVF is a weaker condition than the existence of a KVF since
the geometry and physics are completely unchanged along the integral curves of
\ a Killing vector field (KVF). However, the existence of a non-trivial HVF
leads to restrictions on the equations of state.

Therefore the paper is organized as follows. In section II we outline the
model as well as the self-similar solution. In section III we shall study a
massive cosmic string model with barotropic equation of state under the
self-similar hypothesis. We put special emphasis in calculating the values of
the equation of state which made that the solution follows a power law. In
section IV we shall study a massive cosmic string model with barotropic
equation of state under the self-similar hypothesis but allowing that
\textquotedblleft constants\textquotedblright\ $G$ and $\Lambda$ may vary.
Section V is devoted to study the curvature behaviour of the model taking into
account the obtained solutions in the above sections. We end summarizing the
main results.

\section{The model.}

In synchronous co-ordinates the metric is:%
\begin{align}
ds^{2}  &  =-c^{2}dt^{2}+a^{2}(t)dx^{2}+\left(  b^{2}(t)+K^{2}z^{2}%
a^{2}(t)\right)  dy^{2}+\nonumber\\
&  +2Ka^{2}(t)zdxdy+d^{2}(t)dz^{2}, \label{metric}%
\end{align}
where the metric functions $a(t),b(t),d(t)$ are functions of the time
co-ordinate only and $K\in\mathbb{R}$. The introduction of this constant is
essential since if we set $K=1$, as it is the usual way, then there is not SS
solution for the outlined field equations. In this paper we are interested
only in Bianchi II space-times, hence all metric functions are assumed to be
different and the dimension of the group of isometries acting on the spacelike
hypersurfaces is three.

Once we have defined the metric and we know which are its killing vectors,
then we calculate the four velocity. It must verify, $\mathcal{L}_{\xi_{i}%
}u_{i}=0,$ so we may define the four velocity as follows:%
\begin{equation}
u^{i}=\left(  \frac{1}{c},0,0,0\right)  ,
\end{equation}
in such a way that it is verified, $g(u^{i},u^{i})=-1.$

From the definition of the 4-velocity we find that:
\begin{align*}
\theta &  =u_{\,\,;i}^{i}=\frac{1}{c}\left(  \frac{a^{\prime}}{a}%
+\frac{b^{\prime}}{b}+\frac{d^{\prime}}{d}\right)  =\frac{1}{c}H,\\
H  &  =\sum_{i=1}^{3}H_{i},\\
q  &  =\frac{d}{dt}\left(  \frac{1}{H}\right)  -1,\\
\sigma^{2}  &  =\frac{1}{3c^{2}}\left(  H_{1}^{2}+H_{2}^{2}+H_{3}^{2}%
-H_{1}H_{2}-H_{1}H_{3}-H_{2}H_{3}\right)  .
\end{align*}

From equation $\mathcal{L}_{V}g_{ij}=2g_{ij},$ where $V\in\mathfrak{X}(M),$ we
find the following homothetic vector field for the BII metric (\ref{metric}):%
\begin{equation}
V=t\partial_{t}+\left(  1-t\frac{a^{\prime}}{a}\right)  x\partial_{x}+\left(
1-t\frac{b^{\prime}}{b}\right)  y\partial_{y}+\left(  1-t\frac{d^{\prime}}%
{d}\right)  z\partial_{z}, \label{HO}%
\end{equation}
with the following constrains for the scale factors:
\begin{equation}
a(t)=a_{0}t^{a_{1}},\qquad b(t)=b_{0}t^{a_{2}},\qquad d(t)=d_{0}t^{a_{3}},
\label{scales}%
\end{equation}
with $a_{1},a_{2},a_{3}\in\mathbb{R},$ in such a way, that the constants
$a_{i}$ must verify the following restriction (see \cite{Tony1}):%
\begin{equation}
a_{2}+a_{3}-a_{1}=1. \label{restriction}%
\end{equation}

Taking into account the field equations (FE)%
\begin{equation}
R_{ij}-\frac{1}{2}Rg_{ij}=\frac{8\pi G}{c^{4}}T_{ij}-\Lambda g_{ij},
\end{equation}
where $\Lambda$ is the cosmological constant, and the energy-momentum tensor,
$T_{ij},$ for a cloud of massive strings is given by%
\begin{equation}
T_{i}^{j}=\left(  \rho+p\right)  u^{j}u_{i}+pg_{i}^{j}-\lambda x^{j}x_{i},
\label{strin}%
\end{equation}
where $\rho(t)$ is the rest energy density, i.e. is the rest energy density of
the cloud of strings with particles attached to them ($p-$strings).
$\lambda(t)$ is the string tension density, which may be positive or negative,
$u^{i}$ is the four-velocity for the cloud particles. $x^{i}$ is the
four-vector which represents the strings direction which is the direction of
anisotropy and $\rho=\rho_{p}+\lambda,$ where $\rho_{p}$ denotes the particle
energy density (is the cloud rest energy density), i.e. the string tension
density is connected to the rest energy $\rho$ for a cloud of strings
($p-$strings) with particle attached to them by this relation (see \cite{p1}
and \cite{saha1}).

Since there is no direct evidence of strings in the present-day universe, we
are in general, interested in constructing models of a Universe that evolves
purely from the era dominated by either geometric string or massive strings
and ends up in a particle dominated era with or without remnants of strings.

Moreover the direction of strings satisfy the standard relations:%
\begin{align}
u^{j}u_{i}  &  =-x^{j}x_{i}=-1,\nonumber\\
u^{i}x_{i}  &  =0,\\
x^{i}  &  =\left(  0,a^{-1},0,0\right)  .\nonumber
\end{align}

The Raychaudhuri equation reads%
\begin{equation}
\theta^{\prime}=-\frac{1}{3}\theta^{2}-2\sigma^{2}+R_{ij}u^{i}u^{j},
\end{equation}
with%
\begin{equation}
R_{ij}u^{i}u^{j}=-\frac{1}{2}\rho_{p}.
\end{equation}

It is customary to assume a relation between $\rho$ and $\lambda$ in
accordance with the state equation for strings. The simplest one is a
proportionality relation $\rho=\alpha\lambda,$ where the most usual choices of
the constant $\alpha$ are the following ones (see \cite{saha1}). Geometric
strings. Nambu. $\alpha=1,\rho=\lambda,$ $\rho_{p}=0.$ $p-$strings or
Takabayasi strings%
\begin{equation}
\alpha=1+W, \label{es2}%
\end{equation}
with $W\geq0,$ iff $\rho_{p}=W\lambda.$ Reddy strings.$\alpha=-1.$ A more
general \textquotedblleft barotropic\textquotedblright\ equation of state is
$\rho=\rho\left(  \lambda\right)  ,\rho_{p}=\rho-\lambda.$ In the case of
Takabayasi strings if $W$ is very small, then only geometric strings appear.
On the other hand, if $W$ is infinitely large, then particles dominate over strings.

\section{Barotropic equation of state}

We shall consider the Takabayasi's equation of state, i.e. $\rho=\alpha
\lambda,$ with $\alpha=1+W,$ so the resulting FE are:%

\begin{align}
\frac{a^{\prime}}{a}\frac{b^{\prime}}{b}+\frac{a^{\prime}}{a}\frac{d^{\prime}%
}{d}+\frac{d^{\prime}}{d}\frac{b^{\prime}}{b}-\frac{K^{2}}{4}\frac{a^{2}c^{2}%
}{b^{2}d^{2}}  &  =\frac{8\pi G}{c^{2}}\rho,\\
\frac{b^{\prime\prime}}{b}+\frac{d^{\prime\prime}}{d}+\frac{d^{\prime}}%
{d}\frac{b^{\prime}}{b}-\frac{3K^{2}}{4}\frac{a^{2}c^{2}}{b^{2}d^{2}}  &
=-\frac{8\pi G}{c^{2}}\left(  \omega\rho-\lambda\right)  ,\\
\frac{d^{\prime\prime}}{d}+\frac{a^{\prime\prime}}{a}+\frac{a^{\prime}}%
{a}\frac{d^{\prime}}{d}+\frac{K^{2}}{4}\frac{a^{2}c^{2}}{b^{2}d^{2}}  &
=-\frac{8\pi G}{c^{2}}\omega\rho,\\
\frac{b^{\prime\prime}}{b}+\frac{a^{\prime\prime}}{a}+\frac{a^{\prime}}%
{a}\frac{b^{\prime}}{b}+\frac{K^{2}}{4}\frac{a^{2}c^{2}}{b^{2}d^{2}}  &
=-\frac{8\pi G}{c^{2}}\omega\rho,\\
\rho^{\prime}+\rho\left(  \omega+1\right)  \left(  \frac{a^{\prime}}{a}%
+\frac{b^{\prime}}{b}+\frac{d^{\prime}}{d}\right)  -\lambda\frac{a^{\prime}%
}{a}  &  =0.
\end{align}
noting that $b=d.$ So once again the model collapses to a LRS\ BII\ model (see
\cite{Tony1}) where%
\[
a(t)=a_{0}t^{a_{1}},\qquad b(t)=b_{0}t^{a_{2}}=d(t),
\]
and the constrain $a_{2}+a_{3}-a_{1}=1,$ reads now $2a_{2}-a_{1}=1.$ Therefore
the FE yield:%

\begin{align}
2\frac{a^{\prime}}{a}\frac{b^{\prime}}{b}+\frac{b^{\prime2}}{b^{2}}%
-\frac{K^{2}c^{2}}{4}\frac{a^{2}}{b^{4}}  &  =\frac{8\pi G}{c^{2}}\rho,\\
2\frac{b^{\prime\prime}}{b}+\frac{b^{\prime2}}{b^{2}}-\frac{3K^{2}c^{2}}%
{4}\frac{a^{2}}{b^{4}}  &  =-\frac{8\pi G}{c^{2}}\left(  \omega\rho
-\lambda\right)  ,\\
\frac{b^{\prime\prime}}{b}+\frac{a^{\prime}}{a}\frac{b^{\prime}}{b}%
+\frac{a^{\prime\prime}}{a}+\frac{K^{2}c^{2}}{4}\frac{a^{2}}{b^{4}}  &
=-\frac{8\pi G}{c^{2}}\omega\rho,\\
\rho^{\prime}+\rho\left(  \frac{a^{\prime}}{a}+2\frac{b^{\prime}}{b}\right)
-\lambda\frac{a^{\prime}}{a}  &  =0,
\end{align}

In this way, following the standard procedure (see \cite{Tony1}) we find that%
\[
\rho=\rho_{0}t^{-\beta},\qquad\beta=\left(  4a_{2}-1\right)  \left(
\omega+1\right)  -\frac{\left(  2a_{2}-1\right)  }{1+W},
\]
and hence we need only to solve the system of algebraic equations%
\begin{align}
3a_{2}^{2}-2a_{2}-\frac{3K^{2}}{4}  &  =-A\left(  \omega-\frac{1}{1+W}\right)
,\label{pal1}\\
7a_{2}^{2}-8a_{2}+2+\frac{K^{2}}{4}  &  =-\omega A,\label{pal2}\\
\left(  4a_{2}-1\right)  \left(  \omega+1\right)  -\frac{\left(
2a_{2}-1\right)  }{1+W}  &  =2, \label{pal3}%
\end{align}
with%
\[
A=5a_{2}^{2}-2a_{2}-\frac{K^{2}}{4},\qquad\rho_{0}=\frac{Ac^{2}}{8\pi
G},\qquad\lambda=\frac{\rho}{1+W}.
\]

By solving the system (\ref{pal1}-\ref{pal3}) we obtain the following results:%
\begin{equation}
a_{2}=\frac{2+\omega\left(  1+W\right)  +3W}{2\left(  1+2W+2\omega\left(
1+W\right)  \right)  },
\end{equation}
and%
\begin{equation}
K=\frac{\sqrt{\left(  W-2\right)  ^{2}+2\omega\left(  W^{2}+3W+2\right)
-3\omega^{2}\left(  1+W\right)  ^{2}-8}}{\left(  1+2W+2\omega\left(
1+W\right)  \right)  }.
\end{equation}

We may list the physical solutions for different values of $\omega$

\begin{enumerate}
\item If we set $\omega=0,$ then we get:%
\[
K=\frac{\sqrt{W^{2}-4\left(  W+1\right)  }}{1+2W},\qquad a_{2}=\frac
{2+3W}{2\left(  1+2W\right)  },
\]
and the following restriction: $W>2+2\sqrt{2}.$ Therefore $K\in\left(
0,0.5\right)  $, while $a_{2}\in\left(  0.75,0.77\right)  $ and $a_{1}%
\in\left(  0.5,0.54\right)  .$

\item If $\omega=1/3,$%
\[
K=\frac{\sqrt{-27-24W+12W^{2}}}{5+8W}\Longrightarrow\quad K\in\left(
0,0.43301\right)  ,
\]
with $W>1+\frac{\sqrt{13}}{2}=2.8028,$ therefore the scale factor $a_{2}$
behaves as:%
\[
a_{2}=\frac{7+10W}{2\left(  5+8W\right)  }\quad\Longrightarrow\quad a_{2}%
\in\left(  0.625,0.638\right)  ,
\]
therefore $a_{1}\in\left(  0.25,0.277\right)  .$
\end{enumerate}

In the following table we summarize the behavior of the main quantities for
this model\begin{widetext}
\[%
\begin{array}
[c]{|c|c|c|c|c|c|}\hline
\omega & W & K(W) & a_{2}(W) & a_{1}(W) & OBS\\\hline\hline
1 & <0 &  &  &  & unphysical\\\hline
1/3 & >2.8028 & \left(  0,0.43301\right)  & \left(  0.625,0.638\right)  &
\left(  0.25,0.277\right)  & physical\\\hline
0 & >4.8284 & \left(  0,0.5\right)  & \left(  0.75,0.77\right)  & \left(
0.5,0.54\right)  & physical\\\hline
-1/3 & <0 &  &  &  & unphysical\\\hline
-2/3 & <0 &  &  &  & unphysical\\\hline
\end{array}
\]
\end{widetext}and the energy density behaves as:%
\[
\rho=\rho_{0}t^{-\beta},\quad\beta=\left(  4a_{2}-1\right)  \left(
\omega+1\right)  -\frac{\left(  2a_{2}-1\right)  }{1+W}\approxeq2.
\]

Therefore we are only able to obtain solutions for $\omega=1/3$ and
$\omega=0.$ For the rest of equations of state the obtained solution is
unphysical since in all these cases we have obtained $W<0,$ in order to get
$K\in\mathbb{R}$. But we have ruled out the equation of state $W<0.$

With the obtained results we can see that
\[
H=\frac{4a_{2}-1}{t},\qquad q=\frac{2-4a_{2}}{4a_{2}-1}<0,
\]
note that $2-4a_{2}<0,$ $\forall\omega\in\left[  0,1/3\right]  ,$ while the
shear behaves as:%
\[
\sigma^{2}=\frac{\left(  a_{2}-1\right)  ^{2}}{3c^{2}t^{2}}\rightarrow0.
\]

As it is observed, these quantities fit perfectly with the current
observations obtained by High -Z Supernova Team and Supernova Cosmological
Project (see for example \cite{G},\cite{Per},\cite{Riess} and \cite{Sch}).

\section{Variable constants.}

In this model the resulting FE are:%

\begin{align}
\frac{a^{\prime}}{a}\frac{b^{\prime}}{b}+\frac{a^{\prime}}{a}\frac{d^{\prime}%
}{d}+\frac{d^{\prime}}{d}\frac{b^{\prime}}{b}-\frac{K^{2}}{4}\frac{a^{2}c^{2}%
}{b^{2}d^{2}}  &  =\frac{8\pi G}{c^{2}}\rho+\Lambda c^{2},\label{j1}\\
\frac{b^{\prime\prime}}{b}+\frac{d^{\prime\prime}}{d}+\frac{d^{\prime}}%
{d}\frac{b^{\prime}}{b}-\frac{3K^{2}}{4}\frac{a^{2}c^{2}}{b^{2}d^{2}}  &
=-\frac{8\pi G}{c^{2}}\left(  \omega\rho-\lambda\right)  +\Lambda
c^{2},\label{j2}\\
\frac{d^{\prime\prime}}{d}+\frac{a^{\prime\prime}}{a}+\frac{a^{\prime}}%
{a}\frac{d^{\prime}}{d}+\frac{K^{2}}{4}\frac{a^{2}c^{2}}{b^{2}d^{2}}  &
=-\frac{8\pi G}{c^{2}}\omega\rho+\Lambda c^{2},\label{j3}\\
\frac{b^{\prime\prime}}{b}+\frac{a^{\prime\prime}}{a}+\frac{a^{\prime}}%
{a}\frac{b^{\prime}}{b}+\frac{K^{2}}{4}\frac{a^{2}c^{2}}{b^{2}d^{2}}  &
=-\frac{8\pi G}{c^{2}}\omega\rho+\Lambda c^{2},\label{j4}\\
\rho^{\prime}+\rho\left(  \omega+1\right)  \left(  \frac{a^{\prime}}{a}%
+\frac{b^{\prime}}{b}+\frac{d^{\prime}}{d}\right)   &  =\lambda\frac
{a^{\prime}}{a},\label{j5}\\
\Lambda^{\prime}  &  =-\frac{8\pi G^{\prime}}{c^{4}}\rho. \label{j6}%
\end{align}
noting that $b=d.$ So once again the model collapses to a LRS\ BII\ model (see
\cite{Tony1}), where%
\[
a(t)=a_{0}t^{a_{1}},\qquad b(t)=b_{0}t^{a_{2}}=d(t),
\]
and the constrain $a_{2}+a_{3}-a_{1}=1,$ reads now $2a_{2}-a_{1}=1,$ as above.
Therefore the FE yield:%

\begin{align}
2\frac{a^{\prime}}{a}\frac{b^{\prime}}{b}+\frac{b^{\prime2}}{b^{2}}%
-\frac{K^{2}c^{2}}{4}\frac{a^{2}}{b^{4}}  &  =\frac{8\pi G}{c^{2}}\rho+\Lambda
c^{2},\label{iga01}\\
2\frac{b^{\prime\prime}}{b}+\frac{b^{\prime2}}{b^{2}}-\frac{3K^{2}c^{2}}%
{4}\frac{a^{2}}{b^{4}}  &  =-\frac{8\pi G}{c^{2}}\left(  \omega\rho
-\lambda\right)  +\Lambda c^{2},\label{iga02}\\
\frac{b^{\prime\prime}}{b}+\frac{a^{\prime}}{a}\frac{b^{\prime}}{b}%
+\frac{a^{\prime\prime}}{a}+\frac{K^{2}c^{2}}{4}\frac{a^{2}}{b^{4}}  &
=-\frac{8\pi G}{c^{2}}\omega\rho+\Lambda c^{2},\label{iga03}\\
\rho^{\prime}+\rho\left(  \frac{a^{\prime}}{a}+2\frac{b^{\prime}}{b}\right)
-\lambda\frac{a^{\prime}}{a}  &  =0,\label{iga04}\\
\Lambda^{\prime}  &  =-\frac{8\pi G^{\prime}}{c^{4}}\rho. \label{iga05}%
\end{align}
with
\begin{equation}
\rho=\alpha\lambda,\quad\alpha=1+W. \label{jenny}%
\end{equation}

From eqs. (\ref{iga04}) and (\ref{jenny}) we get%
\begin{equation}
\rho=\rho_{0}t^{-\gamma},\quad\gamma=\left(  4a_{2}-1\right)  \left(
\omega+1\right)  -\frac{\left(  2a_{2}-1\right)  }{1+W}, \label{t_den}%
\end{equation}
where $\gamma=\left(  \omega+1\right)  a-\frac{a_{1}}{1+W}$ and $a=\left(
a_{1}+2a_{2}\right)  $, $2a_{2}-a_{1}=1,$ and we shall assume that $\rho
_{0}>0.$

From eq. (\ref{iga01}) we obtain:%
\begin{equation}
\Lambda=\frac{1}{c^{2}}\left[  At^{-2}-\frac{8\pi G}{c^{2}}\rho_{0}t^{-\gamma
}\right]  , \label{peta1}%
\end{equation}
where $A=2a_{1}a_{2}+a_{2}^{2}-\frac{K^{2}}{4}.$

Now, taking into account eq. (\ref{iga05}) and eq. (\ref{peta1}), algebra
brings us to obtain%
\begin{equation}
G=G_{0}t^{\gamma-2},\qquad G_{0}=\frac{c^{2}A}{4\pi\rho_{0}\gamma}, \label{G}%
\end{equation}
and therefore the cosmological \textquotedblleft constant\textquotedblright%
\ behaves as:
\begin{equation}
\Lambda=\frac{A}{c^{2}}\left(  1-\frac{2}{\gamma}\right)  t^{-2}=\Lambda
_{0}t^{-2}.
\end{equation}

With all these results, we find that the system to solve is the following one:%
\begin{align}
3a_{2}^{2}-2a_{2}-\frac{3K^{2}}{4}  &  =-\frac{2A}{\gamma}\left(  \left(
\omega+1\right)  -\frac{\gamma}{2}-\frac{1}{1+W}\right)  ,\\
7a_{2}^{2}-8a_{2}+2+\frac{K^{2}}{4}  &  =-\frac{2A}{\gamma}\left(  \left(
\omega+1\right)  -\frac{\gamma}{2}\right)  ,
\end{align}
and whose solutions may be listed as follows:

\begin{enumerate}
\item If $\omega=0.$%
\begin{equation}
a_{1}=2a_{2}-1,
\end{equation}
and%
\[
K=-\frac{\sqrt{W\left(  W^{2}-4\left(  W+1\right)  \right)  }}{W\left(
1+2W\right)  }\in\left(  -\frac{1}{2},0\right)  ,
\]
therefore this solution has only sense if $W\geq2+2\sqrt{2}\thickapprox
4.8284.$ In this way we find that%
\[
a_{2}=\frac{2+3W}{2\left(  1+2W\right)  }\in\left(  0.75,0.77\right)  ,
\]
and the behaviour for the rest quantities is:%
\begin{align*}
\rho &  \thickapprox t^{-2},\quad\gamma=\left(  4a_{2}-1\right)
-\frac{\left(  2a_{2}-1\right)  }{1+W}\approxeq2,\\
G  &  \thickapprox t^{\gamma-2}=G_{0},\quad G_{0}=\frac{c^{2}A}{4\pi\rho_{0}%
}\frac{1}{\gamma}=const.>0,\\
\Lambda &  \thickapprox t^{-2},\quad\Lambda_{0}=\frac{A}{c^{2}}\left(
1-\frac{2}{\gamma}\right)  =0,
\end{align*}
with $A=a_{2}\left(  5a_{2}-2\right)  -\frac{K^{2}}{4}>0.$ Note that this
solution coincides with the above one where the constants behaves as true
constants and $\Lambda$ vanishes.

\item If $\omega=1$%
\[
K=-\frac{\sqrt{-27W^{2}-22W-7-12W^{3}+4W^{4}}}{\left(  -1+W\right)  \left(
3+4W\right)  },
\]
so $K\in\left(  -\frac{1}{2},0\right)  $ with $W>4.7,$ and
\[
a_{2}=\frac{-1+3W+6W^{2}}{2\left(  -W-3+4W^{2}\right)  }\in\left(
0.75,0.90\right)  ,
\]
hence $a_{1}\in\left(  0.5,0.80\right)  ,$ with $A=a_{2}\left(  5a_{2}%
-2\right)  -\frac{K^{2}}{4}\thickapprox1.25>0,$ and therefore
\begin{align*}
\rho &  \thickapprox t^{\gamma},\quad\gamma=2\left(  4a_{2}-1\right)
-\frac{\left(  2a_{2}-1\right)  }{1+W}\in\left(  4,5.06\right)  ,\\
G  &  \thickapprox t^{\gamma-2},\qquad G_{0}=\frac{c^{2}A}{4\pi\rho_{0}}%
\frac{1}{\gamma}>0,\\
\Lambda &  \thickapprox t^{-2},\qquad\Lambda_{0}=\frac{A}{c^{2}}\left(
1-\frac{2}{\gamma}\right)  \in\left(  \frac{1}{2},\frac{3}{5}\right)  .
\end{align*}

\item If $\omega=1/3,$ we get the following solution:%
\[
K=-\frac{\sqrt{-519W^{2}-78W-15-456W^{3}+144W^{4}}}{\left(  -1+W\right)
\left(  5+8W\right)  },
\]
so $K\in\left(  -\frac{1}{2},0\right)  ,$ with $W>4.1,$ and
\[
a_{2}=\frac{-1+23W+36W^{2}}{2\left(  7W-5+24W^{2}\right)  }\in\left(
0.75,0.81\right)  ,
\]
hence $a_{1}\in\left(  0.5,0.63\right)  $, therefore
\begin{align*}
\rho &  \thickapprox t^{\gamma},\quad\gamma=\frac{4}{3}\left(  4a_{2}%
-1\right)  -\frac{\left(  2a_{2}-1\right)  }{1+W}\in\left(  2.66,2.86\right)
,\\
G  &  \thickapprox t^{\gamma-2},\qquad G_{0}=\frac{c^{2}A}{4\pi\rho_{0}}%
\frac{1}{\gamma}>0,\\
\Lambda &  \thickapprox t^{-2},\qquad\Lambda_{0}=\frac{A}{c^{2}}\left(
1-\frac{2}{\gamma}\right)  >0.
\end{align*}

\item If $\omega=-1/3,$ then
\[
K=-\frac{\sqrt{-243W^{2}-6W+9-228W^{3}+36W^{4}}}{\left(  1+3W\right)  \left(
1+4W\right)  },
\]
so $K\in\left(  -\frac{1}{2},0\right)  $ with \ $W>6.6,$ and%
\[
a_{2}=\frac{-1+11W+18W^{2}}{2\left(  7W+1+12W^{2}\right)  }\in\left(
0.75,0.7507\right)  ,
\]
hence $a_{1}\in\left(  0.5,0.501\right)  ,$ therefore
\begin{align*}
\rho &  \thickapprox t^{\gamma},\quad\gamma=\frac{2}{3}\left(  4a_{2}%
-1\right)  -\frac{\left(  2a_{2}-1\right)  }{1+W}\in\left(  1.26,1.33\right)
,\\
G  &  \thickapprox t^{\gamma-2},\qquad G_{0}=\frac{c^{2}A}{4\pi\rho_{0}}%
\frac{1}{\gamma}>0,\\
\Lambda &  \thickapprox t^{-2},\qquad\Lambda_{0}=\frac{A}{c^{2}}\left(
1-\frac{2}{\gamma}\right)  <0.
\end{align*}

\item If $\omega=-2/3,$%
\[
K=-\frac{\sqrt{-132W^{2}-48W+48-132W^{3}+9W^{4}}}{\left(  2+3W\right)  \left(
-1+2W\right)  },
\]
in such a way that $K\in\left(  -\frac{1}{2},0\right)  $ with $W>15.2,$ and
\[
a_{2}=\frac{-4+2W+9W^{2}}{2\left(  W-2+6W^{2}\right)  }\in\left(
0.75,0.752\right)  ,
\]
hence $a_{1}\in\left(  0.5,0.504\right)  ,$ therefore
\begin{align*}
\rho &  \thickapprox t^{\gamma},\quad\gamma=\frac{1}{3}\left(  4a_{2}%
-1\right)  -\frac{\left(  2a_{2}-1\right)  }{1+W}\in\left(  0.63,0.66\right)
,\\
G  &  \thickapprox t^{\gamma-2},\qquad G_{0}=\frac{c^{2}A}{4\pi\rho_{0}}%
\frac{1}{\gamma}>0,\\
\Lambda &  \thickapprox t^{-2},\qquad\Lambda_{0}=\frac{A}{c^{2}}\left(
1-\frac{2}{\gamma}\right)  <0.
\end{align*}

\end{enumerate}

In the following table we have summarized the behavior of all the
quantities\begin{widetext}
\[%
\begin{array}
[c]{|c|c|c|c|c|c|c|c|c|}\hline
\omega & W & K(W) & a_{2} & a_{1} & \gamma & G\left(  \gamma-2\right)  &
\Lambda & OBS\\\hline\hline
1 & >4.7 & \left(  -0.5,0\right)  & \left(  0.75,0.90\right)  & \left(
0.5,0.80\right)  & \left(  4,5.06\right)  & \left(  2,3.06\right)  \nearrow &
>0\searrow & phys\\\hline
1/3 & >4.1 & \left(  -0.5,0\right)  & \left(  0.75,0.81\right)  & \left(
0.5,0.63\right)  & \left(  2.6,2.8\right)  & \left(  0.6,0.8\right)  \nearrow
& >0\searrow & phys\\\hline
0 & >4.8 & \left(  -0.5,0\right)  & \left(  0.75,0.77\right)  & \left(
0.5,0.54\right)  & 2 & const. & 0 & phys\\\hline
-1/3 & >6.6 & \left(  -0.5,0\right)  & \left(  0.75,0.7507\right)  & \left(
0.5,0.501\right)  & \left(  1.26,1.33\right)  & \left(  -0.74,-0.67\right)
\searrow & <0\searrow & phys\\\hline
-2/3 & >15.2 & \left(  -0.5,0\right)  & \left(  0.75,0.752\right)  & \left(
0.5,0.504\right)  & \left(  0.63,0.66\right)  & \left(
-1.\,37,-1.\,34\right)  \searrow & <0\searrow & phys\\\hline
\end{array}
\]
\end{widetext}

Therefore, as we can see, and comparing this table with the above one obtained
in the last section we may conclude that to allow that the \textquotedblleft
constants\textquotedblright\ may vary enlarge the set of solutions. In this
case we have obtained a solution for equation of state $\omega\in(-1,1]$ while
in the standard solution we only have a solution for $\omega=0,$ and
$\omega=1/3.$ In the same way it is interesting to emphasize that for
$\omega>0$ i.e. $\omega\in(0,1]$, the \textquotedblleft
constant\textquotedblright, $G$, is a growing time function while the
cosmological \textquotedblleft constant\textquotedblright, $\Lambda$, behaves
as positive decreasing time function (see for example \cite{G},\cite{Per}%
,\cite{Riess} and \cite{Sch}). For $\omega=0,$ we have that both
\textquotedblleft constants\textquotedblright, $G$ and $\Lambda,$ behave as
true constants, in fact we have obtained the same solution as in the above
section. If $\omega<0$ i.e. $\omega\in(-1,0)$, the \textquotedblleft
constant\textquotedblright, $G$, is a decreasing time function while the
cosmological \textquotedblleft constant\textquotedblright, $\Lambda$, behaves
as negative decreasing time function, so, from the observational data we may
rule out these solutions.

With the obtained results we can see that
\[
H=\frac{4a_{2}-1}{t},\qquad q=\frac{2-4a_{2}}{4a_{2}-1}<0,
\]
note that $2-4a_{2}<0,$ and $4a_{2}-1>0,$ $\forall\omega\in\left[  0,1\right]
,$ while the shear behaves as:%
\[
\sigma^{2}=\frac{\left(  a_{2}-1\right)  ^{2}}{3c^{2}t^{2}}\rightarrow0.
\]

As it is observed, these quantities fit perfectly with the current
observations obtained by High -Z Supernova Team and Supernova Cosmological
Project (see for example \cite{G},\cite{Per},\cite{Riess} and \cite{Sch}).

\section{Curvature behaviour.}

With all these results, we find the following behaviour for the curvature
invariants (see for example \cite{Caminati}-\cite{Barrow}).

Ricci Scalar, $I_{0},$ yields%
\begin{equation}
I_{0}=\frac{2}{c^{2}t^{2}}\left(  11a_{2}^{2}-10a_{2}+2-\frac{K^{2}c^{2}}%
{4}\right)  ,
\end{equation}
while Krestchmann scalar, $I_{1}:=R_{ijkl}R^{ijkl}$, yields:%
\begin{align*}
I_{1}  &  =\frac{1}{4c^{4}t^{4}}\left(  432a_{2}^{4}-960a_{2}^{3}+896a_{2}%
^{2}-384a_{2}+64\right. \\
&  \left.  +K^{2}c^{2}\left(  11K^{2}c^{2}-40a_{2}^{2}+80a_{2}-48\right)
\right)  .
\end{align*}

The full contraction of the Ricci tensor, $I_{2}:=R_{ij}R^{ij},$ is:%
\begin{align*}
I_{2}  &  =\frac{1}{4c^{4}t^{4}}\left(  528a_{2}^{4}-1024a_{2}^{3}%
+768a_{2}^{2}-256a_{2}+32\right. \\
&  \left.  K^{2}c^{2}\left(  3K^{2}c^{2}-16a_{2}+8\right)  \right)  ,
\end{align*}
this means that the model is singular.

The non-zero components of the Weyl tensor. The following components of the
Weyl tensor run to $\pm\infty$ when $t\rightarrow0,$%
\[
C_{txtx}\thickapprox C_{txty}\thickapprox t^{4\left(  a_{2}-1\right)  },\quad
C_{tztz}\thickapprox t^{2\left(  a_{2}-1\right)  },
\]
and
\[
C_{tyty}\thickapprox t^{2\left(  a_{2}-1\right)  }+z^{2}t^{4\left(
a_{2}-1\right)  },
\]
these others run to zero when $t\rightarrow0,$%
\[
C_{xzyz}\thickapprox C_{xzxz}\thickapprox C_{xyxy}\thickapprox t^{2\left(
3a_{2}-2\right)  },
\]%
\[
C_{yzyz}\thickapprox t^{2\left(  2a_{2}-1\right)  }+z^{2}t^{2\left(
3a_{2}-2\right)  }.
\]%
\[
C_{txyz}\thickapprox C_{tyxz}\thickapprox C_{tyyz}\thickapprox C_{tzxy}%
\thickapprox t^{4a_{2}-3},
\]
noting that when $a_{2}\rightarrow3/4,$ then they run to a const.

The Weyl scalar, $I_{3}=C^{abcd}C_{abcd}=I_{1}-2I_{2}+\frac{1}{3}I_{0}^{2},$
(this definition is only valid when $n=4)$%
\begin{align*}
I_{3}  &  =\frac{4}{3c^{4}t^{4}}\left(  \left[  \left(  a_{2}-1\right)
\left(  -2\left(  a_{2}-1\right)  -3Kc\right)  +K^{2}c^{2}\right]  \right. \\
&  \left.  \times\left[  \left(  a_{2}-1\right)  \left(  -2\left(
a_{2}-1\right)  +3Kc\right)  +K^{2}c^{2}\right]  \right)  .
\end{align*}

The electric part scalar $I_{4}=E_{ij}E^{ij},$ (see W.C. Lim et al
\cite{Coley})%
\[
I_{4}=\frac{1}{6c^{4}t^{4}}\left(  -2\left(  a_{2}-1\right)  ^{2}+K^{2}%
c^{2}\right)  ^{2},
\]
while the magnetic part scalar $I_{5}=H_{ij}H^{ij},$ yields%
\[
I_{5}=\frac{3}{2c^{2}t^{4}}\left(  a_{2}-1\right)  ^{2}K^{2}.
\]

The Weyl parameter (see W.C. Lim et al \cite{Coley}) which is a dimensionless
measure of the Weyl curvature tensor
\[
\mathcal{W}^{2}=\frac{W^{2}}{H^{4}}=\frac{1}{6H^{4}}\left(  E_{ij}%
E^{ij}+H_{ij}H^{ij}\right)  =\frac{1}{24H^{4}}C^{abcd}C_{abcd},
\]
where%
\[
W^{2}=\frac{1}{36c^{4}t^{4}}\left(  \left(  a_{2}-1\right)  ^{2}+K^{2}%
c^{2}\right)  ^{2}\left(  \left(  2a_{2}-2\right)  ^{2}+K^{2}c^{2}\right)
^{2},
\]
note that the value of $W^{2}$ is really small. We may calculate the quantity
\[
\mathcal{W}^{2}=\frac{\left(  \left(  a_{2}-1\right)  ^{2}+K^{2}c^{2}\right)
^{2}\left(  \left(  2a_{2}-2\right)  ^{2}+K^{2}c^{2}\right)  ^{2}}%
{36c^{4}\left(  4a_{2}-1\right)  ^{4}}\thickapprox const,
\]
but as it is observed $\mathcal{W}\rightarrow0.$ $\mathcal{W}$ can be regarded
as describing the intrinsic anisotropy in the gravitational field. Hence we
may conclude that the model is close to be isotropic. (see \cite{Wain}) since
we have that it is verified the Weyl isotropization criterion i.e.
$\mathcal{W}\rightarrow0$.

The gravitational entropy (see \cite{gron1}-\cite{gron2})
\[
P^{2}=\frac{I_{3}}{I_{2}}=\frac{I_{1}-2I_{2}+\frac{1}{3}I_{0}^{2}}{I_{2}%
}=\frac{I_{1}}{I_{2}}+\frac{1}{3}\frac{I_{0}^{2}}{I_{2}}-2=const.,
\]
since we are working with a SS solution (see \cite{lake} and \cite{Tony2} for
a discussion), and all the dimensionless quantities remain constant along
timelike homothetic trajectories as the Weyl parameter.

\section{Conclusions}

We \ have studied two massive cosmic string Bianchi type II cosmological
models under the self-similar hypothesis. In the first of the studied models
(those with \textquotedblleft constant\textquotedblright\ constants) we have
obtained that the self-similar solution is only valid if the equation of state
is $\omega=0$ and $\omega=1/3.$ For the rest of possible values of $\omega$
the solution is unphysical. In the same way we have shown that such solutions
are only valid if $W$ (the parameter in the equation of state for the strings)
is $W\geq2.8$ (if $\omega=1/3),$ and $W\geq4.8$ (if $\omega=0).$ So may
conclude that the self-similar solution is quite restrictive. Furthermore the
obtained solution collapses to a LRS\ BII solution since two of the scale
factors are equal.

In the second of the studied models (those with \textquotedblleft
varying\textquotedblright\ constants) we have shown that considering a time
varying constants we may enlarge the set of solutions. In this case we have
obtained a mathematical solution for $\omega\in(-1,1],$ and with similar
restrictions for $W.$ In the same way it is interesting to emphasize that for
$\omega>0$ i.e. $\omega\in(0,1]$, the \textquotedblleft
constant\textquotedblright, $G$, is a growing time function while the
cosmological \textquotedblleft constant\textquotedblright, $\Lambda$, behaves
as positive decreasing time function. This fact fits perfectly with the
current observational data. For $\omega=0,$ we have that both
\textquotedblleft constants\textquotedblright, $G$ and $\Lambda,$ behave as
true constants, in fact we have obtained the same solution as in the first of
the studied models. If $\omega<0$ i.e. $\omega\in(-1,0)$, the
\textquotedblleft constant\textquotedblright, $G$, is a decreasing time
function while the cosmological \textquotedblleft constant\textquotedblright,
$\Lambda$, behaves as negative decreasing time function, so, from the
observational data we may rule out these solutions.

\end{document}